\documentclass[aps,pra,twocolumn,showpacs]{revtex4}
\usepackage{graphicx} 
\usepackage{amsmath}
 
\newcommand{\beq}{\begin{equation}}
\newcommand{\enq}{\end{equation}}
\newcommand{\bea}{\begin{eqnarray}}
\newcommand{\ena}{\end{eqnarray}}
\newcommand{\la}{\langle}
\newcommand{\ra}{\rangle}

\begin{document}

\title{Multiband bosons in optical lattices}
\author{J. Larson}
\author{A. Collin}
\author{J.-P. Martikainen}
\email{jpjmarti@nordita.org}
\affiliation{NORDITA, 106 91 Stockholm, Sweden}
\date{\today}

\begin{abstract}
We study a gas of repulsively interacting bosons in an optical lattice and explore
the physics beyond the lowest band Hubbard model.
Utilizing a generalized Gutzwiller ansatz, we find how the lowest band physics is modified by the inclusion of the first excited bands.
In contrast to the prediction of the lowest band Bose-Hubbard model,
a reentrant behavior of superfluidity is envisaged as well as decreasing width of the Mott lobes at strong coupling.
\end{abstract}

\maketitle 

\section{Introduction}
\label{sec:intro}

Experiments with cold atoms in optical lattices have made
dramatic progress in recent past~\cite{Bloch2008a,Lewenstein2007a}.
Optical lattices, low densities, and low temperatures render
fantastic degree of control  and have
enabled detailed studies of strongly correlated quantum systems. For example,
the Mott-superfluid transition~\cite{Jaksch1998a,Greiner2002a} 
has been successfully observed in optical lattices.  This transition
can occur even at $T=0$ and is due to the competition between kinetic energy and
repulsive on-site interactions between lowest band bosons. For large interactions,
the energy is minimized in an incompressible state with fixed
atom numbers while for weaker interactions kinetic energy favors
atomic tunneling and drives the system into a superfluid.

Using Feshbach resonances~\cite{Duine2004a}, or by deepening the lattice, one can increase the strength of the atom-atom interaction compared with the kinetic energy of the atoms in the lattice.
These stronger interactions can cause considerable
excited band populations~\cite{Diener2006a} and take the system
beyond the region where the usual lowest band description is adequate
for a quantitatively accurate description~\cite{Koutsier2006a}. Substantial excited band populations caused by interactions have indeed
been experimentally observed~\cite{Kohl2006a}.
In another interesting experiment, Winkler {\it et al.}~\cite{Winkler2006a}
used Feshbach resonance to increase the interaction energy so much that
the energy of two atom states lied in the bandgap. In this way they
demonstrated the existence of repulsively bound pairs, which were
unable to dissociate due to the absence of available final atomic states.
However, these experiments did not probe the region where the energy
would have been even higher and dissociation to excited bands
would have become possible.

Another way for the excited bands to play a relevant role is to directly couple atoms from the lowest band to the excited bands. 
This was experimentally demonstrated recently by M\"uller {\it et al.}~\cite{Mueller2007a} by coupling atoms from the lowest 
band Mott insulator into the first excited $p$-band of the lattice via Raman transitions between bands. They found lifetimes 
of $p$-band atoms which were considerably longer than the tunneling time-scale in the lattice and were also able to 
explore how coherence on the excited band was established.
This experiment paves the way to explore the equilibrium physics of the purely $p$-band bosons~\cite{Isacsson2005a} 
and outlines a possible route to realize supersolids~\cite{Scarola2005a} or novel phases~\cite{Liu2006a,Xu2007a} on the excited 
bands of an optical lattice. Alternatively, higher bands can be populated by fermions if the filling factor is larger than 
one~\cite{Wu2008a}. In this case the Pauli exclusion principle ensures that the Fermions that cannot populate the lowest band, must 
occupy the excited bands~\cite{Kohl2005a}.  When the optical lattice geometry is varied from the usual
cubic structure, exciting analogs also appear between excited orbital physics and graphene~\cite{CongjunWu2008a}.

In this paper we explore the role of the first excited bands on the phase diagrams of the Bose-Hubbard model. We will 
show that the Mott lobes can become strongly deformed due to combined effects of interactions and excited bands. In some 
parameter regions, reentrant behavior of the insulator superfluid
transitions may arise and in the strongly interacting regime, complex phase diagrams with the possibility of excited 
band superfluidity appears.

The paper is organized as follows. In Sec.~\ref{sec:theory}
we discuss the physical system and derive the Hamiltonian of the system.
Here we also explain the variational approach we use to solve
for the ground state configuration. In Sec.~\ref{sec:results} we proceed by solving for the multiband phase diagrams in different dimensions and discuss the salient features of how excited bands change the lowest band
physics. We also discuss briefly how our results would relate
to results in a trap and what kind of changes excited band atoms
will cause in the trapped system compared with the usual
lowest band results.
We conclude with some discussion in Sec.~\ref{sec:conclusions}.

\section{Theory and and generalized Gutzwiller ansatz}
\label{sec:theory}
The microscopic Hamiltonian for the dilute Bose gas
at low temperatures in a trap is given by
\beq
\begin{array}{lll}
\hat{H}_{microscopic} & = & \displaystyle{\int d{\bf r}\hat\psi^\dagger({\bf r})\left[
-\frac{\hbar^2\nabla^2}{2m}+V({\bf r})
\right]\hat\psi({\bf r})}\\ \\
& & \displaystyle{\!+\!\frac{g}{2}
\hat\psi^\dagger({\bf r})\hat\psi^\dagger({\bf r})
\hat\psi({\bf r})\hat\psi({\bf r})
\!-\!\mu\hat\psi^\dagger({\bf r})\hat\psi({\bf r})},
\end{array}
\label{eq:Hmicroscopic}
\enq
where $\mu$ is the chemical potential, $m$ the atomic mass, $g$ is the interatomic interaction strength, and  $\hat\psi({\bf r})$ and $\hat\psi^\dagger({\bf r})$ are the bosonic annihilation and creation operators, while $V({\bf r})$ is the external trapping potential which in this
work is taken to be the lattice potential
\beq
V({\bf r})=V_L\sum_{\alpha\in \{x,y,z\}} \sin^2\left(
\frac{\pi {\bf r}_\alpha}{d}
\right),
\enq
where $d$ is the lattice spacing and $V_L$ the lattice depth. 
For a deep lattice 
it is reasonable  to expand field operators
in terms of the localized Wannier states. Here we go beyond the
usual lowest band Hubbard model by also including the
first excited states ($p$-band). In a three dimensional lattice
this implies an expansion of the field operators
\beq
\hat\psi({\bf r})=\sum_{{\bf i},\sigma} w_{\sigma,{\bf i}}({\bf r}) 
\hat\psi_{\sigma,{\bf i}},
\enq
where ${\bf i}=(i_x,i_y,i_z)$ labels the lattice site
and $\sigma\in \{0,x,y,z\}$ is the flavor index.
The bosonic operators $\hat\psi_{\sigma,{\bf i}}$ annihilate
a boson of flavor $\sigma$ from the site ${\bf i}$.
We compute the Wannier functions and bandgaps from the ideal gas
band structure.

Substituting the operator expansions into 
Eq.~(\ref{eq:Hmicroscopic}) and ignoring all but the leading
order on-site interactions and nearest neighbor tunneling
processes we derive our fundamental Hamiltonian
\beq
\hat{H}=\hat{H}_0+\hat{H}_{nn}+\hat{H}_{FD},
\enq
where the ideal part is given by
\beq
\hat{H}_0=\sum_{{\bf i}} \left(\Delta_{BG,\sigma}-\mu\right){\hat \psi}_{\sigma,{\bf i}}^\dagger{\hat \psi}_{\sigma,{\bf i}}
-\sum_{\sigma,\alpha}\sum_{<{\bf i},{\bf j}>_\alpha} t_{\alpha,\sigma}
{\hat \psi}_{\sigma,{\bf i}}^\dagger{\hat \psi}_{\sigma,{\bf j}}.
\enq
Here $\Delta_{BG,\sigma}$ are the bandgaps and
$\sum_{<{\bf i},{\bf j}>_\alpha}$ indicates the nearest neighbor sum
over the neighbors in the direction $\alpha\in\{x,y,z\}$.
Since the Bloch functions diagonalize the one-body hamiltonian, there
are no interband hopping terms in the Wannier representation considered
here \cite{Georges2007a}.
The terms originating from the interatomic interactions
are given by
\beq
\hat{H}_{nn}\!=\! \displaystyle{\sum_{\bf i}\sum_\sigma \frac{U_{\sigma\sigma}}{2}
{\hat n}_{\sigma,\bf i}\!\left({\hat n}_{\sigma,\bf i}-1\right)} 
 \displaystyle{+2\!\!\,\sum_{\bf i}\!\sum_{\sigma\sigma',\sigma\neq\sigma'}\!\!
U_{\sigma\sigma'}{\hat n}_{\sigma,\bf i}{\hat n}_{\sigma'\!,\bf i}}
\enq
and
\beq
\hat{H}_{FD}\!=\!\displaystyle{\sum_{\bf i}\!\!\sum_{\sigma\sigma',\sigma\neq\sigma'} 
\!\!\frac{U_{\sigma\sigma'}}{2}}\!\left(\!{\hat \psi}_{\sigma,{\bf i}}^\dagger {\hat \psi}_{\sigma,{\bf i}}^\dagger {\hat \psi}_{\sigma',{\bf i}} {\hat \psi}_{\sigma',{\bf i}}
+{\hat \psi}_{\sigma',{\bf i}}^\dagger {\hat \psi}_{\sigma',{\bf i}}^\dagger {\hat \psi}_{\sigma,{\bf i}}{\hat \psi}_{\sigma,{\bf i}}
\!\right),
\enq
where $\hat{H}_{FD}$ contains terms that describe flavor changing
collisions and collisions transferring atoms between bands. This term has formal 
similarity with terms responsible
for spin-dynamics in spinor condensates~\cite{Law1998a,Stamper-Kurn1998b}. 
However, the strength
of these terms is comparable to other interaction terms
as opposed to spinor condensates where it is usually
small, being proportional to the difference between singlet
and triplet scattering lengths (for spin-$1$ spinor condensate).
If the lattice site were precisely harmonic there could also be processes
where two atoms on the $p$-band collide and scatter into an atom
on the lowest band and an atom on the $d$-band. 
In a real lattice, such processes
are off-resonant and they are also ignored in our model restricted
to just the lowest bands. For the lattice depth considered in 
this paper, the anharmonicity is indeed evident, and only 
momentum states close to the Brillouin edges are non far off-resonant. 
In addition, the population of $p$-band bosons is in general 
small and the corresponding scattering amplitudes 
have been verified to be typically one order of 
magnitude smaller than for non-flavor changing collisions.

Further, terms describing scattering between atoms in 
neighboring sites have  been left out as well. 
For our choice of lattice depth, we have checked that 
the magnitude of these particle assisted tunneling terms are 
1-2 $\%$ of the $s$-band on-site scattering amplitude, and hence 
such an approximation is justified. It should be kept in mind though, that for 
other system parameters there are circumstances when nearest neighbor 
interactions~\cite{Scarola2005a} or
particle assisted tunneling processes~\cite{Duan2008a} might give 
rise to new physics, but these regimes are not considered here.

The various coupling strengths in the lattice model are
related to $g$ through
\beq
U_{\sigma\sigma'}=g\int d{\bf r} w_\sigma({\bf r})^2w_{\sigma'}({\bf r})^2
\enq
and the tunneling coefficients are given by
\beq
t_{\sigma,\alpha}=-\int d{\bf r} w_\sigma({\bf r})
\left[-\frac{\hbar^2\nabla^2}{2m}+V({\bf r})\right]
w_\sigma({\bf r}-d{\bf e}_\alpha),
\enq
where ${\bf e}_\alpha$ is the unit vector in the direction $\alpha$.
When the lattice is symmetric, the tunneling strength on the lowest
band is independent of direction. However, the directional
dependence of the tunneling strength must be kept
for the $p$-band atoms, since the overlap integrals are very different
depending on whether one is integrating along the node of the
Wannier function or orthogonal to it. This can have important
consequences for the characteristics in these 
systems~\cite{Isacsson2005a,Martikainen2008a}

The above formulation was derived for the three-dimensional
situation with $4$ flavors. In this paper we will, however,
also consider the corresponding special cases of one- and two-dimensional systems with $2$ and $3$ flavors respectively. The number of relevant
flavors can be reduced in  asymmetric lattices,
where the bandgap in the direction of deep lattice potential
becomes higher than in other directions.

We approach the physics of the above theory using a Gutzwiller
ansatz for the many body wave-function generalized to 
multiple flavors~\cite{Buonsante2008a}. Our ansatz is given by
\beq
|\psi\ra=\prod_{\bf i} \sum_{\bf n }f_{{\bf n}}^{({\bf i})} |{\bf n}\ra_{\bf i},
\enq
where $f_{{\bf n}}^{({\bf i})}$ are the variational
parameters to be determined by minimizing the energy.
In our Gutzwiller approach,
the on-site states are expanded in terms of the Fock
states $|{\bf n}\ra=|n_0,n_x,n_y,n_z\ra$
of the multiple flavor system. For numerical reasons
the sum over ${\bf n}$ must be cut-off and in our computations
we include all the states with a total excitation
$\sum_\sigma n_\sigma$ less than $8$. The number of variational
parameters in a site increase proportional to $n_{cutoff}^{N_f}$ where
$N_f$ is the number of flavors. For this reason, the problem
is substantially heavier in terms of computing power than implementing
the Gutzwiller ansatz for a single flavor problem.

For the lowest band Bose-Hubbard model it is known
that the Gutzwiller method provides 
a reasonably accurate description of the transitions 
from Mott-insulators in three- and two-dimensions.
In a more strongly correlated 
one-dimensional system it is reasonable only qualitatively
with considerable quantitative differences from exact computations.
This gives us confidence that the above ansatz is able
to capture most of the interesting physics expected to occur 
in our model.

We minimize the energy functional at $T=0$ using a conjugate
gradient method. Since the signs of the tunneling strengths
can vary and flavor changing collisions are sensitive to phase
factors in the wave-function amplitudes,
we have observed that one has to be 
careful in ensuring a convergence to the real lowest energy state
as opposed to some local minima. Among other things, 
negative tunneling strength
in the kinetic energy can favor $\pi$-phase modulation
of the condensate order parameter and if this modulation
is not present in the initial guess input, the numerics might not converge
to the real minimum. Flavor dynamics is also sensitive to the 
phase factors between flavors and it is easy to converge to solutions with
incorrect relative phase factors between different flavor order parameters.

\section{Results}
\label{sec:results}
\subsection{One-dimensional lattice with two flavors}
\label{sec:results1D}

\begin{figure}
\includegraphics[width=0.90\columnwidth]{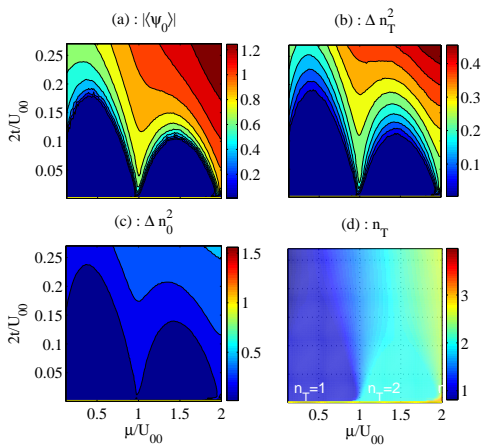}
\caption{(Color online)
Properties of the one-dimensional two flavor Bose-Hubbard model
as a function of chemical potential and the inverse 
interaction strength $2t/U_{00}$, where 2 is the number of nearest neighbors.
For concreteness the parameters were computed for a lattice
of depth $V_L=15\, E_R$. (a) Lowest band condensate order parameter,
(b) total atom number fluctuation, (c) fluctuation of the lowest 
band atom number, and (d) the total on-site atom number.
 The horizontal yellow line indicates
the position $t/\Delta_{BG}$ and warm (cold) colors imply high (low) values.
}
\label{fig:1DPhaseDiagram}
\end{figure}

\begin{figure}
\includegraphics[width=0.90\columnwidth]{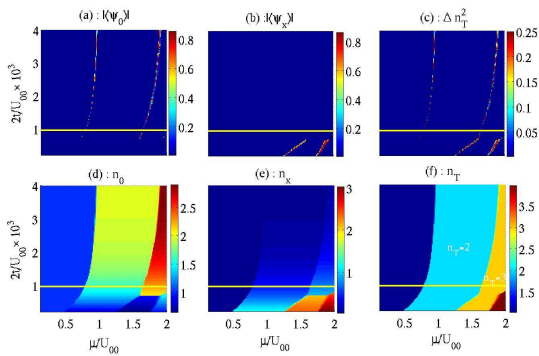} 
\caption{(Color online)
Properties of the one-dimensional two flavor Bose-Hubbard model
as a function of chemical potential and the inverse 
interaction strength $2t/U_{00}$ at strong coupling.
(a)-(b) Lowest and excited band condensate order parameters,
(c) total atom number fluctuation, (d)-(e) lowest and excited
band atom numbers, and (f) the total on-site atom number.
 The horizontal yellow line indicates
the position $t/\Delta_{BG}\times 10^3$ and warm (cold) 
colors imply high (low) values.
Lattice parameters are the same as in the Fig.~\ref{fig:1DPhaseDiagram}.
}
\label{fig:1DPhaseDiagram_Zoom}
\end{figure}

In a one-dimensional lattice our model reduces to the lowest
band and one excited $p$-band. We summarize our results in
Figs.~\ref{fig:1DPhaseDiagram} and ~\ref{fig:1DPhaseDiagram_Zoom}. Throughout, the figures display the order parameters $\langle\psi_\alpha\rangle$, 
on-site total and lowest band number fluctuations $\Delta n_T^2$ 
and $\Delta n_0^2$, and  on-site total number and flavor number of 
atoms $n_T$ and $n_\alpha$. These quantities are presented as functions 
of the scaled 
hopping parameter $zt/U_{00}$ (where $z$ is the number of nearest 
neighbours and $t$ the $s$-band tunneling coefficient) and the scaled 
chemical potential $\mu/U_{00}$. The lattice depth for $^{87}{\rm Rb}$
atoms is fixed at $V_L=15E_R$,
 where $E_R$ is the recoil energy.

When there is only one atom per site, this
atom resides on the lowest band and, for relatively weak coupling,
also for higher atom numbers the result is essentially
the same as for the lowest band Bose-Hubbard model. As expected, one finds
a superfluid region (on the lowest band) at weak coupling
and Mott insulating lobes of different integer atom
number at somewhat stronger coupling. 

However, at  stronger
coupling the population of the excited band increases
smoothly when the on-site atom number 
$n_T=2$. The density difference between flavors
starts to fluctuate and the Mott-lobes become distorted
as they curve toward smaller chemical potentials, giving rise
to re-entrant behavior not present in the lowest band Hubbard model
within the Gutzwiller approximation.
As one moves towards weaker coupling in this region, one can start
from an insulator, move into a superfluid, then  back into an insulator
and finally re-enter the superfluid region.
Note, however, that re-entrant behavior could exist in one-dimensional
models which move beyond the Gutzwiller approximation~\cite{Koller2006a}.
In that case re-entrant behavior is due to correlation effects between sites
and is not caused by the presence of excited bands.

At even stronger couplings, the population on the $p$-band becomes more dominant and 
influences the state of the system more strongly. 
This is depicted in Fig.~\ref{fig:1DPhaseDiagram_Zoom}, 
which zooms on the regime of small $t/U_{00}$. In the multiflavor problem Mott regions can change also
qualitatively. In particular, one can often have insulating
regions where the total on-site atom number is non-fluctuating, but
where the densities of individual flavors, or their density differences 
do fluctuate. This is possible even in the limit of infinitely deep lattice
due to flavor changing on-site collisions. Physics on the excited band is
quite sensitive to the details of the model, such as the relative magnitude 
of the bandgap and the lowest band bandwidth. With the parameter values
than those used here, there is a possibility of narrow regions of excited band
superfluidity appearing in the phase diagram.

Interestingly there is a peculiar transition
around $\mu/U_{00}\approx 1.5$ at strong coupling. As one
goes across this region in the direction of weaker coupling
one moves from a Mott-insulator with $n_T=3$ predominantly
on the excited band, into a narrow excited band
superfluid region, and finally into a Mott-insulator
 with $n_T=3$ mainly on the lowest band. Across this transition the flavor
densities $\la n_\sigma\ra$ change abruptly.
The difference in the Mott-insulators
around this transition is that at strong coupling
the state is predominantly of type $\alpha|0,3\ra+\beta|2,1\ra$ while
at weaker coupling it is of type $\alpha|3,0\ra+\beta|1,2\ra$.
Close to the transition (in the narrow SF region) there are 
more non-zero amplitudes in the wave-function expansion.

In a mean-field Gross-Pitaevskii picture one could think
of this transition as somewhat analogous to a transition in a system
of two-component Bose-Einstein condensates. When there is a strong repulsion
between components,
it is more favorable to put more atoms into the state with a 
maximum density near the edges and a vanishing density in the center.
At weaker, coupling it is more favorable to have more atoms
in the "core'' state. Related transitions are predicted by 
the one-dimensional multi-orbital mean-field computations~\cite{Alon2005a}.

\subsection{Two-dimensional lattice with three flavors}
\label{sec:results2D}
\begin{figure}
\includegraphics[width=0.90\columnwidth]{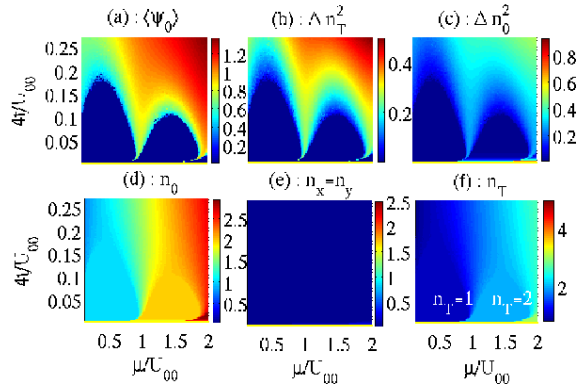} 
\caption{(Color online)
Properties of the two-dimensional three flavor Bose-Hubbard model
as a function of chemical potential and the inverse 
interaction strength $4t/U_{00}$.
The parameters are computed for a lattice
of depth $V_L=15\, E_R$. In (a) we show a contour plot of the lowest
band condensate order parameter. In (b) and (c) we show a contour plot of 
the total number fluctuation and the lowest band number fluctuation, respectively. Finally, (d) and (e) show a surface plot of the individual flavor densities and (f) the total density. The horizontal yellow line indicates the position $t/\Delta_{BG}$ 
and warm (cold) colors imply high (low) values.
}
\label{fig:2DPhaseDiagram}
\end{figure}

\begin{figure}
\includegraphics[width=0.90\columnwidth]{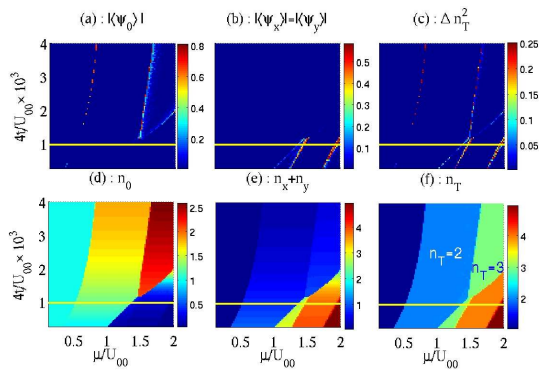} 
\caption{(Color online) 
Properties of the two-dimensional three flavor Bose-Hubbard model
as a function of chemical potential and the inverse 
interaction strength $4t/U_{00}$ at strong coupling.
In (a)-(c) we show the lowest and excited band order parameters
as well as the total number fluctuation.
In (d)-(f) we show the lowest and excited band populations together
with the total on-site atom number. 
The horizontal yellow line indicates
the position $10^3\times t/\Delta_{BG}$ 
and warm (cold) colors imply high (low) values.
The lattice parameters are the same as in Fig.~\ref{fig:2DPhaseDiagram}.
}
\label{fig:2DPhaseDiagram_Zoom}
\end{figure}

In a two-dimensional symmetric lattice there are two-degenerate
$p$-bands which must be taken into account.
Our results for the two dimensional system are shown
in  Figs.~\ref{fig:2DPhaseDiagram} and ~\ref{fig:2DPhaseDiagram_Zoom}, 
where ~\ref{fig:2DPhaseDiagram_Zoom} 
gives a close-up in the regime of small $t/U_{00}$.
As can be seen from the figures, the Mott lobes are deformed by the
excited bands in a similar way as in a one-dimensional system with fewer
flavors. At stronger coupling the lowest band atom number can 
start to fluctuate when the on-site atom number is greater than one and 
this is caused by the possibility of the excited band carrying 
part of the site occupation.

In the regime of strong coupling, the expected phase diagram
is quite complex with the possibility of transitions between lowest band and 
excited band superfluids. Also, around $\mu/U_{00}\approx 1.2$ at strong coupling there is a transition (as one changes $t/U_{00}$) between
a Mott insulator with $n_T=3$ to a Mott insulator with $n_T=2$. Here 
the population of the lowest band is reduced abruptly as one moves
towards strong coupling. Around $\mu/U_{00}\approx 1.5$, a similar transition is encountered between insulators with $n_T=3$ and $n_T=4$ 
with a narrow superfluid region on the excited
bands in between the insulating phases. 
The magnitude of the excited band condensate
order parameters were found to be equal. 

\subsection{Three-dimensional lattice with four flavors}
\label{sec:results3D}
\begin{figure}
\includegraphics[width=0.90\columnwidth]{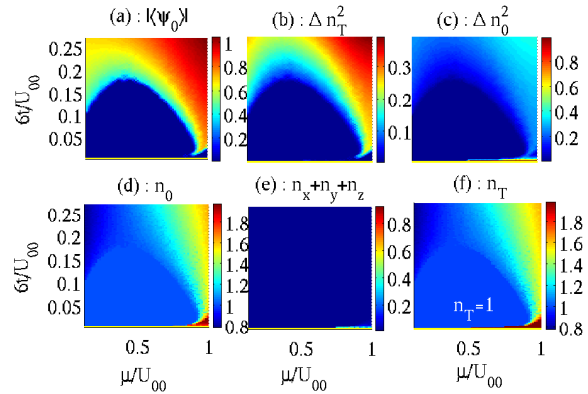}
\caption{(Color online)
Properties of the three-dimensional four flavor Bose-Hubbard model
as a function of chemical potential and the inverse 
interaction strength $6t/U_{00}$.
The parameters are computed for a lattice
of depth $V_L=15\, E_R$. In (a)-(c) we show the 
lowest band condensate order parameter,
total number fluctuation and the number fluctuation of the lowest band atoms. 
Figures (d)-(f) show the lowest band atom number, excited 
band atom number and the total atom number respectively. The horizontal line indicates
the position $t/\Delta_{BG}$ and warm (cold) colors imply high (low) values.
}
\label{fig:3DPhaseDiagram}
\end{figure}

In a three-dimensional lattice where
the lattice depth is equal in all directions, 
the degeneracy of the $p$-band is increased
to $3$ and the total number of relevant flavors is $4$. As the dimensionality
increases the computations become substantially more time-consuming
partly due to larger state space, but also because of more stringent
convergence properties to the global energy minima for large couplings.
However, as we demonstrate in Fig.~\ref{fig:3DPhaseDiagram} the general
structure of the phase diagrams remains for couplings weaker or
of same order of magnitude than the bandgaps. The re-entrant behavior
of the superfluid-insulator transition is therefore independent of
dimensionality. 

For three-dimensional $p$-band superfluids there
is the interesting possibility of frustration of the relative phases
of the condensate order parameters~\cite{Isacsson2005a}. In this paper we have
assumed that the system is homogeneous, but it would be interesting, although
computationally more challenging, to
investigate effects of possible phase frustration in an inhomogeneous system
at very strong coupling.

\subsection{Trapped physics}
In the current contribution we have focused on the physics 
of homogeneous systems, but some qualitative conclusions
about the trapped case are easy enough to draw in the spirit
of a local density approximation in which the chemical potential
is replaced by the local chemical potential
$\mu_{eff}=\mu-V({\bf r})$, where $V({\bf r})$ is the trapping potential.
For the usual harmonic trap this effective chemical potential
is maximized in the center of the trap and is reduced as one
moves to the edge of the cloud. 

With respect to our phase diagrams
this implies that as one moves from the center to the edge
of the cloud, one moves from right to left on our $x$-axis $\mu/U_0$
for some particular value of $zt/U_0$, where $z$ is number of nearest neighbors. In the lowest band
Bose-Hubbard model this suggest either a superfluid everywhere
or a wedding cake phase structure (at stronger coupling), with
Mott-insulating plateaus with different 
atom numbers layered in between superfluid regions. 

Inspecting our phase diagrams one can see that this picture
changes with the inclusion of the excited bands
in a sense that one can see gradually increasing
excited band population as the coupling grows. 
As one increases the coupling strength the width of
the insulating regions first grows, just like
in the lowest band Bose-Hubbard model, but eventually
starts to decrease, in contrast to the predictions of the lowest
band Hubbard model.
At even stronger couplings, a superfluidity on the excited band might
be established before moving to a fully insulating phase as
coupling approaches infinity.

\section{Discussion and concluding remarks}
\label{sec:conclusions}
Having found effects due to excited bands in the context of the bosonic Hubbard model, from an experimental point of view, one would naturally wish to measure the physical properties of the system in a way which is sensitive to the band index. In the pioneering experiments by Greiner {\it et al.}~\cite{Greiner2002a} the Mott insulating regime was observed by vanishing interference peaks after time of flight expansion for deep enough lattices. In our model part of the systems population lies on the excited bands which are typically not phase coherent. After a time of flight such contributions from the excited bands would show up as incoherent background whose weight depends on the fraction of atoms on the excited bands. Furthermore, the detailed structure of the interference peaks depends on the initial Wannier state prior to expansion and the presence of nodes on the Wannier function could, in principle, be observed by the careful observation of the structure of the interference peaks. However, this might be complicated in practice especially in the multi-dimensional systems. Addressing the role of flavor dynamics and number fluctuation in different flavors requires probes which address different flavors individually. This has been demonstrated recently~\cite{Mueller2007a} by coupling atoms with Raman transitions between the lowest band and some excited band. In Ref.~\cite{Mueller2007a} the selectivity was achieved by breaking the degeneracy of $p$-band atoms by lattice anisotropy, but alternatively one could engineer a vanishing transition matrix elements for undesired transitions. Interestingly, Mueller {\it et al.}~\cite{Mueller2007a} were also able to observe time-evolution
of the Bose gas caused by flavor dynamics.

An additional practical challenge is that most clear deviations from
the simple lowest band model occur in the regime where the tunneling strength is small compared to the interaction strengths. While the strong interactions will rapidly reach equilibrium in each individual site, the establishment of global phase coherence between sites would require a longer time since its time-scale is related to the tunneling energy. In order to have time to establish global phase coherence the onsite loss processes should remain sufficiently weak. However, if the interest is primarily on on-site properties such as flavor dynamics and number fluctuations between flavors, then the requirements on the strengths of loss processes are not as stringent since the relevant time-scale is set by the onsite interaction energies.

Here we solved the multiband Bose-Hubbard model
utilizing a generalized Gutzwiller ansatz. Having outlined
how the couplings to the excited band can change
the physics from the lowest band Bose-Hubbard model, we now proceed
to briefly discuss limitations of our formalism.
We cut-off our theory 
to include only the lowest band and the first excited
band(s) of the lattice. This is expected to be a good
approximation when the coupling is clearly smaller than the bandgaps.
However, some of our results are computed in a more
strongly coupled region and in such case one can expect
a non-negligible level populations on even higher bands~\cite{Diener2006a}
and because of this, part of our results might change quantitatively
although not necessarily qualitatively, when more bands
are included in the theory. However, it is unclear
how the details of the phase diagrams in the very strongly coupled
region change if more bands are included in the theory.
Our main interest has been to elucidate how the lowest band results are changed
by the inclusion of more bands and we have found how the Mott lobes
change as the couplings become comparable to bandgaps. This observation
is quite general and not strongly model dependent. However, details
of the excited band physics are expected to be more sensitive
to the parameters and models used. 

With strong interactions and more than one atom per site, the localized 
Wannier functions can be dressed by the interactions. Repulsive interactions would
spread the on-site wave function and therefore lower the
on-site interaction energy and increase the tunneling strength
from those values used in this paper by computing the ideal gas Wannier
functions. Such corrections can give rise to quantitative 
corrections~\cite{Li2006a}
although such corrections are not expected to be
extremely large for lattices with relatively few atoms
in each site.
Here we have not specified how the on-site interaction
$U_{00}$ is to be related to, for example, the three dimensional 
scattering length $a$. We left this relation unspecified
partly because such relation can depend strongly on dimensionality
and partly because in the region of very strong coupling
the two-body scattering physics might change due to
confinement effects. In particular, if one were to estimate
the three-dimensional 
on-site interaction using the usual relation
$U_{00}=g\int d{\bf r} |w_0({\bf r})|^4$ together
with the harmonic approximation one can show that
$
U_{00}/\Delta_{BG}=\sqrt{2/\pi}\left(a/\sigma\right),
$
where $\sigma$ is the size of the Wannier wave-function. This shows
that interaction becomes comparable to bandgaps in the regime
where scattering length is comparable to the size of the localized
wave function and consequently the two-body scattering problem might have
to be considered in the presence of the external 
potential~\cite{Fedichev2004a}.
If one does the similar computation 
in a two-dimensional pancake system of thickness $\sigma_z$, one finds
$U_{00}/\Delta_{BG}\sim a/\sigma_z$ and the coupling becomes
comparable to the bandgap only in the regime of two-dimensional
scattering theory~\cite{Petrov2000b}.

In this paper we have considered the usual single color
optical lattice and we found that rather strong interactions are required
to induce substantial changes in the phase diagram predicted by the
lowest band Bose-Hubbard model.
However, it should be realized that many of the phenomena
we discuss here are expected to play a role also
in a more complicated setting of
two-color superlattices~\cite{Trotzky2008a,Stojanovic2008a}. 
In such systems one can have
a lattice of "deep'' sites which in turn are effectively
double well potentials. In that case the on-site wave function on the lowest
band is the symmetric wave function which peaks in both sites
of the deep site. The first excited wave function is
the antisymmetric and is separated from
the lowest state by an energy proportional to the tunneling
strength in the double well system. This bandgap
is tunable and can be used to bring the bandgap down even when
the lattice of deep sites remains well described by the tight
binding Hubbard model. 

In such system one can expect
more dramatic changes in the phase diagram at lower interaction strengths.
Furthermore, in superlattices
one could engineer band structures (in a one-dimensional system)
with only two bands close in energy, while the next band is substantially
separated from the two lowest bands. Hence, under such circumstances the theory
can be safely cut-off to just include the two lowest bands when
interactions and bandgap between the lowest and the first excited band 
are smaller than the bandgap between
first and second excited bands. However, the physics might not always 
be entirely
similar to that studied in this paper
since density assisted tunneling processes and nearest neighbor
interactions can sometimes 
play a larger role in superlattices~\cite{Trotzky2008a,Barmettler2008a}.


\begin{acknowledgements}
While writing this paper we
became aware that Hui Zhai {\it et al.} have been
working independently on some aspects 
of a similar problem~\cite{ZhaiCastuComment2008a}.
\end{acknowledgements}


\end{document}